\documentclass[11pt]{article}
\usepackage{color}
\usepackage{amssymb}
\usepackage{amsmath}
\usepackage{epsfig}

\definecolor{pink}{rgb}{1,.4,.7}
\definecolor{magenta}{rgb}{1,0,1}
\definecolor{violet}{rgb}{.9,.25,.6}
\definecolor{darkolivegreen3}{rgb}{.6,.8,.35}
\definecolor{maroon3}{rgb}{.8,.26,.56}
\definecolor{mediumorchid}{rgb}{.73,.33,.83}
\definecolor{mediumorchid1}{rgb}{1.,.33,.63}
\definecolor{darkgreen}{rgb}{0.1,.6,.13}
\definecolor{lightyellow}{rgb}{1.,1.,.82}
\definecolor{turquoise}{rgb}{.35,.80,.71}
\definecolor{coral}{rgb}{1.,.6,.21}
\definecolor{orangered}{rgb}{1.,.5,0.}
\definecolor{orange}{rgb}{1.,.65,.1}
\definecolor{blue1}{rgb}{.48,.53,1.}
\definecolor{gold}{rgb}{1.,.85,0.}
\definecolor{darkviolet}{rgb}{.54,.04,.84}

\def\phiq{{\phi}_q}
\def\phix{{\phi}_x}

\def\Journal#1#2#3#4{{#1} {\bf #2}, #3 #4}

\def\etal{{\it et al.}}

\def\APJ{\em ApJ.}
\def\APP{\em Astropart. Phys.}
\def\AST{\em Astron. J.}

\def\GRG{\em Gen. Rel. Grav}
\def\IMA{{\em Int. J. Mod. Phys.} A}
\def\IMP{\em Int. J. Mod. Phys.}
\def\JHE{\em J. High Ener. Phys.}

\def\JPL{\em JETPhys. Lett.}

\def\MPL{{\em Mod. Phys. Lett.} A}

\def\NAT{\em Nature}

\def\NPB{{\em Nucl. Phys.} B}
\def\PLB{{\em Phys. Lett.}  B}
\def\PRL{\em Phys. Rev. Lett.}
\def\PRD{{\em Phys. Rev.} D}

\def\PRE{\em Phys. Rep.}

\def\RPP{\em Rept. Prog. Phys.}

\def\be{\begin{equation}}
\def\ee{\end{equation}}
\def\bea{\begin{eqnarray}}
\def\eea{\end{eqnarray}}

\begin{document}
\begin{center}
\Large \bf {Quintessence From The Decay of a Superheavy Dark Matter}\\
\end{center}

\begin{center}
{\it Houri Ziaeepour\\
{Mullard Space Science Laboratory,\\Holmbury St. Mary, Dorking, Surrey
RH5 6NT, UK.\\
Email: {\tt hz@mssl.ucl.ac.uk}}
}
\end{center}

\medskip
\begin {tabular}{p{12cm}p{3cm}}
 & \bf \it {To Memory of Changoul and her Love}
\end {tabular}
\medskip

\begin {abstract}
We investigate the possibility of replacing the cosmological constant with
gradual condensation of a scalar field produced during the decay of a 
superheavy dark matter.
The advantage of this class of models to the ordinary quintessence is that
the evolution of the dark energy and the dark energy are correlated and 
cosmological
coincidence problem is solved. This model does not need a special form for 
the quitessence potential and even a simple ${\phi}^4$ theory or an axion 
like scalar is enough to explain the existence of the Dark Energy.
We show that the model has an intrinsic feedback between energy density of
the dark matter and the scalar field such that for a large volume 
of the parameter space the equation of state of the scalar field from very
early in the history of the Universe is very close to a cosmological
constant. Other aspects of this model are consistent with recent CMB and
LSS observations.
\end {abstract}

\section {Introduction}
Quintessence models are alternatives to a Cosmological Constant i.e. a 
non-zero vacuum energy density. They are not however flawless. 
Even in models with tracking solutions the potential of the scalar field 
must somehow be fine-tuned to explain its smallness and its slow variation 
until today. In addition, many of them can not address the coincidence 
problem i.e. why the density of Dark Matter (DM) and Dark Energy (DE) 
evolve in such a way that they become comparable just after galaxy formation.

Recently a number of authors have proposed interaction between 
dark matter and quintessence field to explain the coincidence. 
L.P. Chimento \etal ~\cite {interac0} based on an earlier work by L.P. 
Chimento \etal ~\cite {interac1} and W. Zimdahl \etal ~\cite {interac2} 
suggest an asymptotic scaling law between density of DE and DM. In their 
model due to a 
dissipative interaction between dark matter and quintessence scalar field 
${\phi}_q$, 
${\rho}_{dm} / {\rho}_q \rightarrow cte.$ where ${\rho}_{dm}$ and 
${\rho}_q$ are respectively DM and scalar field density. Assuming 
this ``strong coincidence'' ~\cite {interac0}, they find the class of 
potentials $V_q ({\phi}_q)$ such that the equation of state
have a solution with scaling behavior. Then, using constraints from 
nucleosynthesis, they find that this category of models have 
$w_q \gtrsim -0.7$. 
This value is marginally compatible with WMAP data and far from publicly 
available SN-Ia data which prefers $w_q \sim -1$. In another version of the 
same model, W. Zimdahl \etal ~\cite {interac2} consider a non-static scaling 
solution, ${\rho}_{dm} / {\rho}_q \propto (a_0/a)^\eta$. The model 
with $\eta = 1$ 
solves the coincidence paradigm but the standard $\Lambda$CDM fits 
the SN-Ia data better and their best fit has $w_q \sim -0.7$.

L. Amendola \etal ~\cite {interac3} have extensively studied the interaction 
of quintessence field and dark matter in models with tracking solutions 
and $w_q > -1$. They show that these models 
are equivalent to a Brans-Dicke Lagrangian with power law potential and 
look like a ``Fifth Force''. Modification of the CMB 
anisotropy spectra by such interactions is observable and put stringent 
constraints on their parameters. 

D. Comelli \etal ~\cite {massvar} study a model in which the effect of 
interaction between 
quintessence scalar and dark matter appears as time dependence of DM 
particles mass. This explains the extreme adjustment of dark matter and 
dark energy 
densities during cosmological evolution. The coupling between two fields 
increases the parameter space for both and reduces by orders of 
magnitudes the amount of fine tuning. In this respect, as we will see below, 
their model is similar to what we propose in this work. However, there are a 
number of issues that these authors have not addressed. Cosmological 
observations 
put strict limits on the variation of fundamental parameters including the 
DM mass. In their model the largest amount of variation happens around and 
after matter domination epoch. The mass variation must leave an imprint on 
the CMB and large structure formation which was not observed.

In addition to the lack of explanation for coincidence in many quintessence 
models, it is difficult to find a scalar field with necessary 
characteristics in the frame of known particle physics models without some 
fine tuning of the potential ~\cite {quinpot}. In general, it is assumed 
that quintessence 
field is axion with high-order, thus non-renormalizable, interactions with 
the Standard Model particles (or its super-symmetric extension) which is 
highly suppressed at low energies. However, D. Chung \etal ~\cite {massrenor}
show that any supergravity induced interaction between ${\phi}_q$ and other 
scalars with VEV of the order of Plank mass can increase the very tiny 
mass of the ${\phi}_q$ ($m_q \sim H_0 \sim 10^{-33} eV$) expected in many 
models, unless a discrete global symmetry prevents their contribution to the 
mass. 

In a very recent work, G.R. Farrar \& P.J. Peebles ~\cite {yukawa} study 
models with a 
Yukawa interaction between DM and quintessence scalar field. Like 
D. Comelli \etal model, this interaction affects the mass of the dark matter 
particles. The general 
behavior of these models is close to $\Lambda$CDM with some differences 
which can distinguish them. One of the special cases with a 
2-component CDM imitates the $\Lambda$CDM very closely. Many aspects 
of this model is similar to the model studied in the present work but without 
considering the source of the intimate relation between DM and DE in contrast 
(we believe) to the present work. Moreover, the necessity of having a very 
special self-interaction potential for the quintessence field is not removed.

What we propose here is a model for dark energy somehow different from 
previous quintessence models (A preliminary investigation of this model has 
been presented in ~\cite {th2002}). We assume that DE is the result of the 
condensation of a scalar field produced during very slow decay of a massive 
particle. In most of quintessence models the scalar field is produced during 
inflation or reheating period in large amount such that to control its 
contribution to the total energy of the Universe, its potential must be a 
negative exponential (in most cases sum of two exponentials) or a negative 
power function ~\cite {quinpot}. We show that in the present model very 
small production rate of the scalar field replaces 
the fine tuning of the potential and practically any scalar field even 
without a self-interaction has a tracking solution for a large part of its  
parameter space.

The main motivation for this class of models is the possibility of a 
top-down 
solution ~\cite {xpart}~\cite {wimpzilla}~\cite {grapro} for the mystery of 
Ultra High Energy Cosmic Rays (UHECRs) ~\cite 
{fly}~\cite{agasa}~\cite{uhecrrev}. If a very small part of the decay 
remnants which make the primaries of UHECRs is composed of a scalar 
field $\phiq$, its 
condensation can have all the characteristics of a quintessence field. We 
show that in this model the most natural equation of state for the 
quintessence scalar is very close to a cosmological constant, at least until 
the age of the Universe is much smaller than the life-time of the 
Superheavy  
Dark Matter (SDM, WIMPZILLA) which is the origin of the quintessence field. 

Another motivation is the fact that a dark energy with 
$w_q \lesssim -1$ fits the SN-Ia data 
better than a cosmological constant ~\cite {snmeasur}~\cite{snmeasur1}
~\cite {snhouri}. Although the sensitivity of CMB data to the equation of 
state of the dark energy is much less than SNs, with $95\%$ confidence 
WMAP data gives the range 
$-1 \pm 0.22$ for the $w_q$ ~\cite {cmbwmap}~\cite {wq}. Estimation from 
galaxy clusters evolution is also in agreement with this range 
~\cite {clustw}. 
On the other hand, it has been demonstrated that the cosmological equation 
of state for a decaying dark matter in presence of 
a cosmological constant is similar to a quintessence with $w_q \lesssim -1$ 
~\cite {snhouri}. Both observations therefore seem to encourage a 
top-down solution which explains simultaneously the dark energy and the 
UHECRs. 

Like other models with interaction between DM 
and DE, the coincidence in this model is solved without fine-tuning. 
Parameters can be changed by many orders of magnitude without destroying 
the general behavior of the equation of state or the extreme relation 
between the energy density of dark energy and the total energy density in 
the early Universe.

In Sec.\ref {sec.model} we solve the evolution equations for dark matter and 
dark energy. For two asymptotic regime we find analytical solutions for the 
evolution of $\phiq$ . In Sec.\ref {sec.simul} we present the results of 
numerical solution of the evolution equations including the baryonic matter 
and we show that both approaches lead essentially to the same conclusion. We 
study also the extent of the parameter space. The effect of DM anisotropy 
on the energy density of the dark energy is studied in 
Sec.\ref {sec.perturb}. We show that the perturbation of dark energy in this 
model is very small and very far from the resolution of present or near 
future observations. The late time decoherence of the scalar field is 
discussed in Sec.\ref{subsec.decohere}. We give a qualitative 
estimation of the 
necessary conditions and leave a proper investigation of this 
issue as well as the possible candidates for $\phiq$ to future works.

\section {Cosmological Evolution of a Decaying Dark Matter and a 
Quintessence Field} \label {sec.model}
Consider that at very early epoch in the history of the Universe, just after 
inflation, the cosmological ``soup'' 
consists of 2 species: a superheavy dark matter (SDM) - $X$ particles - 
decoupled from the rest of the ``soup'' since very early time and a second 
component which we don't consider in detail. The only constraint we need 
is that it must consist of light species including baryons, 
neutrinos, photons, and light dark matter (by light we mean with respect to 
X). For simplicity we assume that $X$ is a scalar field $\phix$. Considering 
$\phix$ to be a spinor or vector does not change the general conclusions 
of this work. We also assume that $\phix$ is quasi-stable i.e. its lifetime 
is much longer than the present age of the Universe. A very small part of its 
decay remnants is considered to be a scalar field $\phiq$ with negligibly 
weak interaction with other fields.

The effective Lagrangian can be written as:
\be
{\mathcal L} = \int d^4 x \sqrt{-g} \biggl [\frac {1}{2} g^{\mu\nu} 
{\partial}_{\mu} \phix {\partial}_{\nu} \phix + \frac {1}{2} g^{\mu\nu} 
{\partial}_{\mu} \phiq {\partial}_{\nu} \phiq - V (\phix, \phiq, J) 
\biggr ] + {\mathcal L}_J \label {lagrange}
\ee
The field $J$ presents collectively other fields. 
The term $V (\phix, \phiq, J)$ 
includes all interactions including self-interaction potential for 
$\phix$ and $\phiq$:
\be
V (\phix, \phiq, J) = V_q (\phiq) + V_x (\phix) + g {\phix}^m {\phiq}^n + 
W (\phix, \phiq, J) \label {potv}
\ee
The term $g {\phix}^m {\phiq}^n$ is important because it is responsible for 
annihilation of $X$ and back reaction of quintessence field by reproducing 
them. $W (\phix, \phiq, J)$ presents 
interactions which contribute to the decay of $X$ to light fields and to 
$\phiq$ (in addition to what is shown explicitly in 
(\ref{potv})). The very long lifetime of $X$ 
constrains this term and $g$. They must be strongly suppressed. 
For $n = 2$ and $m = 2$ the $g$ term contributes to the 
mass of $\phix$ and $\phiq$. Because of the huge mass of $\phix$ (which 
must come from another coupling) and its very small occupation number 
$<{\phix}^2> \sim 2 {\rho}_x / {m_x}^2$, for sufficiently small $g$ the 
effect of this term on the mass of the SDM is very small. 
We discuss the r\^ole of this term in detail later. If the interaction of 
other fields with $\phiq$ is only through the exchange of $X$ (for instance 
due to a conserved symmetry shared by both $X$ and $\phiq$), the huge mass 
of $X$ suppresses the interaction and therefore the modification of their 
mass. This solves the problem of ``Fifth Force'' in the 
dark ~\cite {interac3} and the SM sectors. 

In a homogeneous universe the evolution equations for $\phiq$ and $\phix$ are:
\bea
\ddot{\phiq} + 3H \dot{\phiq} + \frac {\partial V}{{\partial} \phiq} & = & 0
\label{phiqevo}\\
\ddot{\phix} + 3H \dot{\phix} + \frac {\partial V}{{\partial} \phix} & = & 0
\label{phixevo}
\eea
where dot means the comoving time derivative. In the rest of this work we 
treat 
$\phix$ and $J$ as classical particles and 
deal only with their density and equation of state. We assume that $X$ 
particles are non-relativistic (i.e. part of the CDM) with negligible 
self-interaction i.e.
\be
V_x (\phix) = \frac {1}{2} {m_x}^2 {\phix}^2 \label {vx}
\ee
Under this assumptions $\phix$ can be replaced by:
\be
\phix \sim \biggl (\frac {2 {\rho}_x}{{m_x}^2}\biggr )^{\frac {1}{2}} 
\label {rhox}
\ee
If $X$ is a spinor, the lowest order (Yukawa) interaction term in 
(\ref {lagrange}) is $g \phiq \bar {\psi} \psi$. In the classical treatment 
of $X$:
\be
\bar {\psi} \psi \sim \frac {{\rho}_x}{m_x} \label {spinor}
\ee
The same argument about the negligible effect of the interaction on the mass 
of DM and SM particles is applied. For simplicity we consider only 
the scalar case.

For potential $V_q (\phiq)$ we consider a simple ${\phi}^4$ model:
\be
V_q (\phiq) = \frac {1}{2} {m_q}^2 {\phiq}^2 + \frac {\lambda}{4} {\phiq}^4 \label {vq}
\ee
Conservation of energy-momentum, Einstein and dynamic equations, 
give following system of equations for the fields:
\bea
\dot{\phiq} [\ddot{\phiq} + 3H \dot{\phiq} + {m_q}^2 \phiq + 
\lambda {\phiq}^3] & = & -2g \dot{\phiq}\phiq 
\biggl (\frac {2 {\rho}_x}{{m_x}^2}\biggr ) + {\Gamma}_q{\rho}_x 
\label {phiqe} \\
\dot {{\rho}_x} + 3H {\rho}_x & = & - ({\Gamma}_q + {\Gamma}_J)
{{\rho}_x} - {\pi}^4 g^2 \biggl (\frac {{{\rho}_x}^2}{{m_x}^3} - 
\frac {{{\rho}_q}'^2}{{m_q}^3}\biggr ) \label {xeq} \\
\dot {{\rho}_J} + 3H ({\rho}_J + P_J) & = & {\Gamma}_J {{\rho}_x} 
\label {jeq} \\
H^2 & \equiv & \biggl (\frac {\dot{a}}{a}\biggr )^2 = \frac {8\pi G}{3} 
({\rho}_x + {\rho}_J + {\rho}_q) \label {heq} \\
{\rho}_q & = & \frac {1}{2} {m_q}^2 {\dot{\phiq}}^2 + 
\frac {1}{2} {m_q}^2 {\phiq}^2 + \frac {\lambda}{4} {\phiq}^4 
\label {phidens}
\eea
where (\ref {xeq}) is the Boltzmann equation for $X$ particles. We 
calculate its 
right hand side in the appendix. ${{\rho}_q}'$ is the 
density of quintessence particles (not the classical field $\phiq$) 
with an average energy larger than $m_x$ in the local inertial frame. Only 
interaction between these 
particles contribute to the reproduction of SDM. ${\Gamma}_q$ and 
${\Gamma}_J$ are respectively the decay width of $X$ to $\phiq$ and to other 
species. In (\ref {phiqe}) we have replaces $\phix$ with its classical 
approximation from (\ref {rhox}). 
The effect of decay Lagrangian $W (\phix, \phiq, J)$ appears as 
$({\Gamma}_q + {\Gamma}_J){\rho}_x$ which is the decay rate of $X$ 
particles (see equation (\ref {momcons}) in the Appendix).

At very high temperatures when ${\rho}_x \gg {\pi}^4g^2{m_x}^3 \Gamma$, the 
annihilation and reproduction terms in (\ref {xeq}) are dominant. $X$ 
particles however are non-relativistic up to temperatures close to their rest 
mass. Quintessence scalar particles at this time are relativistic and 
therefore their density falls faster than SDM density by a factor of $a (t)$. 
The probability of annihilation also decreases very rapidly. Consequently, 
from very early time only the decay term in (\ref {xeq}) is important. The 
dominance of annihilation/reproduction can happen only if the production 
temperature of $X$ particles i.e. preheating/reheating temperature is very 
high. Such scenarios however can make dangerous amount of gravitinos ~\cite 
{gravprod}. For this reason, presumably the reheating temperature must be 
much smaller 
than $m_x$ and annihilation dominance never happens. This can not put the 
production of SDM in danger because it has been shown ~\cite {wimprod} that 
even with a very low reheating temperature they can be produced. It seems 
therefore reasonable to study the evolution of the fields only when the 
annihilation/reproduction is negligible. Another reason for this 
simplification 
is that we are interested in the decohered modes of $\phiq$. When the 
self-annihilation of $X$ particles is the dominant source of $\phiq$ most 
of particles are highly relativistic and their self interaction doesn't have 
time to make long-wavelength modes. This claim needs however a detail 
investigation of the process of decoherence which we leave for another work.

The system of equations (\ref {phiqe})-(\ref {phidens}) is highly non-linear 
and an analytical solution 
can not be found easily. There are however two asymptotic regimes which 
permit an approximate analytical treatment. The first one happens at very 
early time just after the production of $X$ (presumably 
after preheating ~\cite {wimpzilla}~\cite {grapro}) and the decoherence of 
$\phiq$'s long wavelength modes. In this epoch  $\phiq \sim 0$ and can be 
neglected. The other regime is when comoving time 
variation of $\phiq$ is very slow and one can 
neglect $\ddot {\phiq}$. We show that the first regime leads to 
a saturation (tracking) solution where $\phiq \rightarrow cte.$ It then can 
be treated as the initial condition for the second regime when $\phiq$ 
changes slowly.

The effect of last term in right hand side of (\ref {xeq}) as we argued 
is negligible. The solution of (\ref {xeq}) is then straightforward:
\be
{\rho}_x (t) = {\rho}_x (t_0) e^{-\Gamma (t-t_0)} \biggl (\frac {a (t_0)}
{a (t)}\biggr )^3 \label {xsol}
\ee
where $\Gamma \equiv {\Gamma}_q + {\Gamma}_J$ is the total decay 
width of $X$. 
We consider $t_0$ to be the time after production and decoupling of $X$.
These two times can be very different, but with an extremely long 
lifetime for $X$ and its weak interaction with other species, it is not 
important which one of them is selected as $t_0$. 

After inserting the solution (\ref {xsol}) and neglecting all the terms 
proportional to $\phiq$, equation (\ref {phiqe}) simplifies to:
\be
\dot{\phiq} \frac {d}{dt}(a^3 \dot{\phiq}) = {\Gamma}_q a^3 (t_0) 
{\rho}_x (t_0) e^{-\Gamma (t-t_0)} \label {phiapp}
\ee
and can be solved:
\be
\frac {1}{2}{\dot{\phiq}}^2 (t) \equiv K_q(t) = \biggl (\frac {a (t_0)}
{a (t)}\biggr )^6 \biggl [K_q(t_0) + {\Gamma}_q {\rho}_x (t_0) 
\int_{t_0}^t dt \frac {a^3}{a (t_0)}e^{-\Gamma (t-t_0)} \biggr ]
\label {phiappsol}
\ee
For $a \propto t^k$ the integral term in (\ref {phiappsol}) decreases with 
time (i.e. $\ddot{\phiq} < 0$). This means that after a relatively short 
time $\phiq$ is saturated and its density does not change, in other words 
it behaves like a 
cosmological constant. The numerical simulation in the next section confirms 
this result. If $\phiq$ was a classical field the natural choice for the 
initial value of the kinetic energy $K_q(t_0)$ was $K_q(t_0) = 0$ assuming 
a very rapid production of $X$. However, in reality $\phiq$ is a 
quantum field and it gets time to decohere and to settle as a classical 
field. The initial value of $K_q(t_0)$ can therefore be non-zero. Its exact 
value can only be determined by investigating the process of decoherence. 
In any case with the expansion of the Universe, its effect on $\dot{\phiq}$ 
decreases very rapidly because of $a^{-6}(t)$ factor in (\ref {phiappsol}).

Next we consider the regime where $\phiq$ changes very slowly and we can 
neglect $\ddot{\phiq}$ and higher orders of $\dot{\phiq}$. Equation 
(\ref{phiqe}) gets the following simplified form:
\be
\dot{\phiq}({m_q}^2 \phiq + \lambda {\phiq}^3) = -2g 
\dot{\phiq}{\phiq} \biggl (\frac {2 {\rho}_x}{{m_x}^2}\biggr ) + 
{\Gamma}_q{\rho}_x \label {phisloworg}
\ee
We expect that self-interaction of $\phiq$ be much stronger than its coupling to $X$. Neglecting the first term in the right hand side of (\ref{phisloworg}), its 
$\phiq$-dependent part can be integrated:
\be
\frac {d}{dt} \biggl (\frac {1}{2} {m_q}^2 {\phiq}^2 + \frac {\lambda}{4}
{\phiq}^4 \biggr ) = \frac {dV}{dt}(\phiq) = {\Gamma}_q {\rho}_x 
\label {phislowphi}
\ee
which then is easily solved:
\be
V_q (\phiq) = V_q (\phiq (t'_0)) + {\Gamma}_q {\rho}_x (t'_0) 
\int_{t'_0}^t dt \biggl (\frac {a (t'_0)}{a (t)}\biggr )^3 
e^{-\Gamma (t-t'_0)} 
\label {phislowsol}
\ee
Here $V_q$ is the potential energy of $\phiq$. From (\ref {phislowphi}) and 
(\ref {phislowsol}) it is clear that the final value of the potential and 
therefore $\phiq$ energy density is driven by the decay term and not the 
self-interaction. Therefore the only vital condition for this model is the 
existence of a long life SDM and not the potential of $\phiq$. 

In (\ref {phislowsol}) the initial 
values $t'_0$ and $\phiq (t'_0)$ are different from equation 
(\ref {phiappsol}). They correspond to the time and to the value of $\phiq$ 
in the first regime when it approaches to saturation. Similar to 
(\ref {phiappsol}), the time 
dependence of $\phiq$ in (\ref {phislowsol}) vanishes exponentially and the 
behavior of $\phiq$ approaches to a cosmological constant. 

To estimate the asymptotic value of $\phiq$ we assume that $a (t) \propto 
t^k$. Using (\ref {phislowsol}) with the additional assumption that 
$t_s - t'_0 \ll 1/\Gamma$, ($t_s$ is the saturation time), we find:
\be
V (\phiq) - V (\phiq (t'_0)) \sim \frac {{\Gamma}_q{\rho}_x (t'_0)}
{(3k -1)}\biggl (1- \biggl (\frac {t'_0}{t} \biggr)^{(3k -1)} \biggr). 
\label {vtime}
\ee
If we define the saturation time as the time when $V (\phiq) - 
V (\phiq (t'_0))$ has 90\% of its final value, for $t_s \ll t_{eq}$ with 
$t_{eq}$ the matter-radiation equilibrium time, $k = 1/2$ and:
\be
t_s \sim 100 t'_0   \label {tsrad}
\ee
For $t_s \gg t_{eq}$, $k = 2/3$ and:
\be
t_s \sim 10 t'_0    \label {tsdm}
\ee
The interesting conclusion one can make from (\ref{vtime}) is that the 
initial density of SDM, its production time, and its decay rate to $\phiq$ 
which are apparently independent quantities determine together the final 
value of the dark energy density. The long lifetime of SDM is expected to be 
due to a symmetry which is broken only by non-renormalizable high order 
weak coupling operators. They become important only at very large energy 
scales. These conditions are exactly what is needed to have a small dark 
energy density according to (\ref{vtime}). In Sec.\ref{sec.simul} we 
see that numerical calculation confirm these results. 

We can also observe here the main difference between this model and other 
quintessence 
models. If $\phiq$ is produced during e.g. the decay of inflaton or from the 
decay of a short live particle in the early Universe, its final density 
should be much 
larger than observed dark energy unless either its production width was 
fine-tuned to unnaturally small values or its self-interaction was 
exponentially suppressed with some fine tuning of its rate.

\subsection {Decoherence} \label {subsec.decohere}
Decoherence of scalar fields has been mainly studied in the context of 
phase transition ~\cite {phasetr} in a thermal system. Examples are phase 
transition in condense matter ~\cite {phasetr}~\cite {decoherethermal}, 
and before, during and after inflation in the early 
Universe ~\cite {decohereinf0}~\cite {decohereinf1}. In the latter case the aim is studying the inflation itself, 
production of defects and the reheating. Decoherence is the result of 
self-interaction as well as interaction between a field (regarded as order 
parameter after decoherence) and other fields in the environment. Long 
wavelength modes behave like a classical field i.e. don't show 
``particle-like'' behavior if quantum correlation between modes are 
negligible. More technically this happens when the density matrix for these 
modes is approximately diagonal. It has been shown ~\cite {decohereinf0} that 
interaction with higher modes is enough to decohere long wavelength modes 
(see Calzetta \etal ~\cite {decohereinf1} for a review). The classical order 
parameter corresponds to these modes after their decoherence. 
One can consider a cut-off in the mode space which separate the 
system (i.e. long wavelength modes) from environment (short wavelengths). 
The cutoff can be considered as an evolving scale which determines at each 
cosmological epoch the decoherent/coherent modes ~\cite{decohereinf1}.

It has been shown ~\cite {decoherethermal} that the decoherence time in a 
thermal phase 
transition is shorter than the spinodal time i.e. the time after beginning 
of the phase transition when the scalar field or more precisely $<{\phi}^2>$ 
settles at the minimum of the potential. The decoherence time in presence 
of external fields (with couplings of the same order as self-interaction) is 
\be
t_d \sim \frac {1}{m} \label {td}
\ee
By replacing Minkovski time with conformal time and considering a time 
dependent cut-off  ~\cite {decohereinf0}~\cite {decohereinf1}
~\cite {decoherethermal}one can show that modes with:
\be
\frac {k^2}{a^2} + m^2 \lesssim H^2 \label {decohere}
\ee
decohere and behave like a classical scalar field. The effect of coupling 
constant is logarithmic and less important.

If the SDM exists, it is produced during preheating ~\cite {grapro} just 
after the end of the inflation presumably at $T \sim 10^{14}eV-10^{16}eV$ 
which correspond to:
\be
H \sim 10^{-6}eV-10^{-4}eV    \label {usize}
\ee
From (\ref{tsrad}) this time 
range permits scalars with mass $m \lesssim 10^{-6}eV$ to decohere. When the 
size of the Universe get larger, $\phiq$ stops decohering. This also helps 
having a very small dark energy density. If the preheating/reheating had 
happened 
when the Hubble constant was smaller, them $m_q$ also must be smaller to have 
long wavelength modes which can decohere. We will see in the next section 
that in this case the main term in $V (\phiq)$ potential is the 
self-interaction. Moreover, $\lambda$ can be larger which helps a faster 
decoherence of long wavelength modes.

The argument given here is evidently very qualitative and needs much deeper 
investigation. In the present work we take the possibility of decoherence as 
granted and study the evolution of $\phiq$ as a classical scalar field.

\section {Numerical Solution} \label {sec.simul}
To have a better understanding of the behavior and the parameter space 
of this model, we have solved 
equations (\ref{phiqe}) to (\ref{phidens}) numerically. We have also added 
the interaction between various species of the Standard Model particles 
to the simulation to be closer to real cosmological evolution and to 
obtain the equation of state of the remnants. This 
is specially important for constraining the lifetime of SDM ~\cite 
{wimpzhouri}. Without considering the interaction between high energy 
remnants and the rest of the SM particles specially the CMB, the lifetime 
of SDM must be orders of magnitude larger than present age of the Universe.

Details of interaction simulation are discussed in ~\cite {wimpzhouri} 
and we don't repeat them here. The Boltzmann equation for SM species 
(equation (1) in ~\cite {wimpzhouri}) replaces (\ref{jeq}). Because of 
numerical limitations we switch on interactions only from $z = 10^9$ 
downward. For the same reason we 
had to begin the simulation of $X$ decay from $z \sim 10^{14}$ which is 
equivalent to a temperature of $T = 10^{11} eV$. The expected reheating 
temperature is model dependent and varies from $\sim 10^{22} eV$ to 
$\sim 10^7 eV$. For the time being no observational constraint on this large 
range is available. The change in the initial temperature however does not 
modify the results of the simulation significantly if 
$f_q \equiv {\Gamma}_q/{\Gamma}$ is rescaled inversely proportional to 
redshift and to the 
total decay width ${\Gamma}$, and proportional to $m_x$. In other words two 
models lead to very similar results for the quintessence field if:
\be
\frac {f_q}{f'_q} = \frac {z' {\Gamma}' m_x}{z {\Gamma} m'_x} \label {rescal}
\ee

\begin{figure}[h]
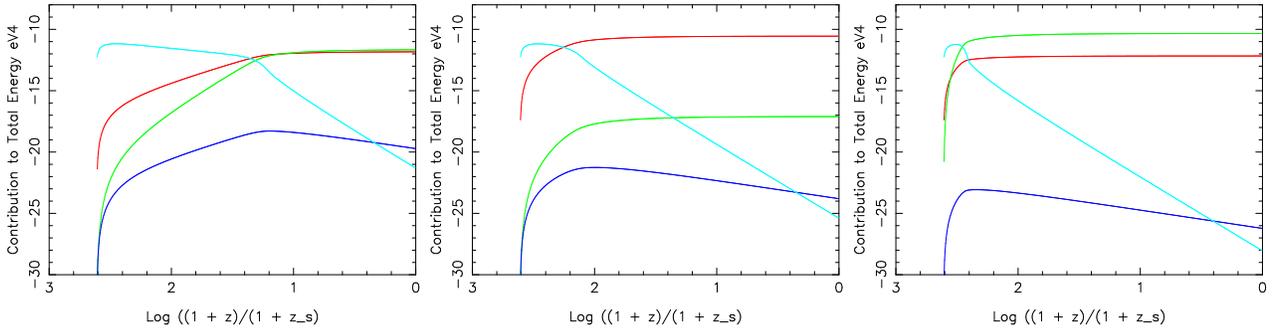

\begin{center}
\psfig{figure=quincontribmass-8.eps,angle=-90,width=5.5cm}
\psfig{figure=quincontrib.eps,angle=-90,width=5.5cm}
\psfig{figure=quincontriblambda-10.eps,angle=-90,width=5.5cm}
\caption{Evolution of the contribution to the total energy density of $\phiq$ 
for ${\Gamma}_0 \equiv {\Gamma}_q/\Gamma = 10^{-16}$ and : Left, 
$m_q = 10^{-8} eV$ and $\lambda = 10^{-20}$; Center, $m_q = 10^{-6} eV$ 
and $\lambda = 10^{-20}$; Right, $m_q = 10^{-6} eV$ and $\lambda = 
10^{-10}$.Curves are: mass (red), self-interaction (green), 
kinetic energy (cyan) and interaction with SDM (blue).
\label {fig:quincontrib}}
\end{center}
\end{figure}

For the lifetime of 
$X$ we use the results from ~\cite {wimpzhouri} and ~\cite {snhouri} which 
show that a lifetime $\tau = 5 {\tau}_0 - 50 {\tau}_0$ (${\tau}_0$ the 
present 
age of the Universe) can explain the observed flux of UHECRs as well as 
cosmic equation of state with $w_q \lesssim -1$. In the following we 
consider $\tau = 5 {\tau}_0$. Our test shows that increasing $\tau$ to 
$50 {\tau}_0$ does not significantly modifies the extent of the admissible 
parameter space or other main characteristics of the dark energy model.
We consider only the models with $m = 2$, $n = 2$ and $g = 10^{-15}$ in 
(\ref{potv}). The results for $10^{-20} \leqslant g \leqslant 10^{-5}$ are 
roughly the same as what we present in this section and therefore they are 
not shown. The discussion in Sec.\ref{sec.model} as well as 
Fig.\ref {fig:quincontrib} show that the contribution of the interaction 
with the SDM in the total energy density of $\phiq$ is much smaller than 
other terms.

\begin{figure}[h]
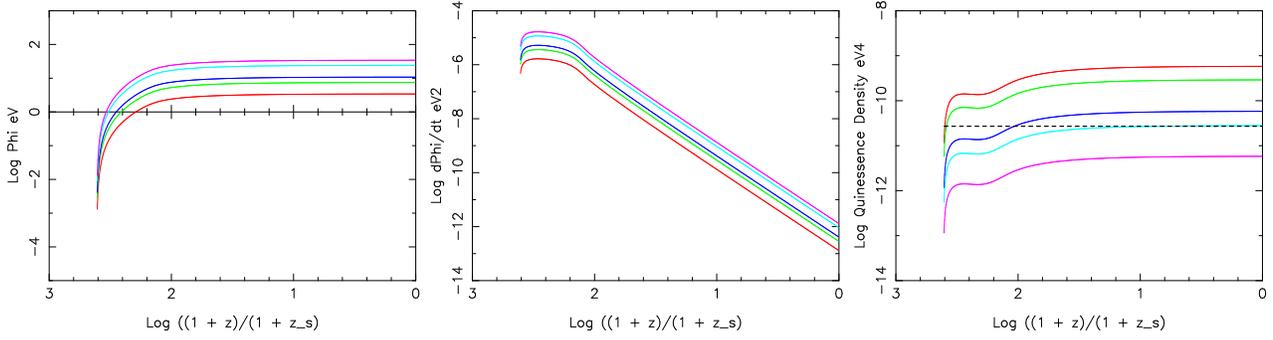

\begin{center}
\psfig{figure=quinphi.eps,angle=-90,width=5.5cm}
\psfig{figure=quindphi.eps,angle=-90,width=5.5cm}
\psfig{figure=quindens.eps,angle=-90,width=5.5cm}
\caption{Evolution of quintessence field (left), its derivative (center) 
and its total energy density (right) for ${\Gamma}_0 \equiv 
{\Gamma}_q/\Gamma = 10^{-16}$ 
(magenta) (see text for details), $5 {\Gamma}_0$ (cyan), $10 {\Gamma}_0$ 
(blue), $50 {\Gamma}_0$ (green), $100 {\Gamma}_0$ (red). Dash line is the 
observed value of the dark energy. $m_q = 10^{-6} eV$, 
$\lambda = 10^{-20}$.
\label {fig:quinevol}}
\end{center}
\end{figure}
Fig.\ref{fig:quinevol} shows the evolution of $\phiq$, its time derivative 
and its total energy density from the end of $X$ production to saturation 
redshift $z_s$. Here we have used as $z_s$ the redshift after which up to 
simulation precision the total energy density of $\phiq$ does not change 
anymore. The result of the simulation is quite consistent with the 
approximate solutions discussed in Sec.\ref{sec.model}. The final density 
energy of $\phiq$ is practically proportional to ${\Gamma}_q/\Gamma$. 
The latter quantity encompasses 3 important parameters of the model: The 
fraction of 
energy of the remnants which changes to $\phiq$, the fraction of energy in 
the long wavelength modes which can decohere and the coupling of these 
modes to the environment which contributes to $\phiq$ yield and to the 
effective formation redshift of the classical quintessence field $\phiq$. 
Therefore the effective volume of the parameter space presented by this 
simulation is much larger and the fine-tuning of parameters are much less 
than what is expected from just one parameter.

\begin{figure}[h]
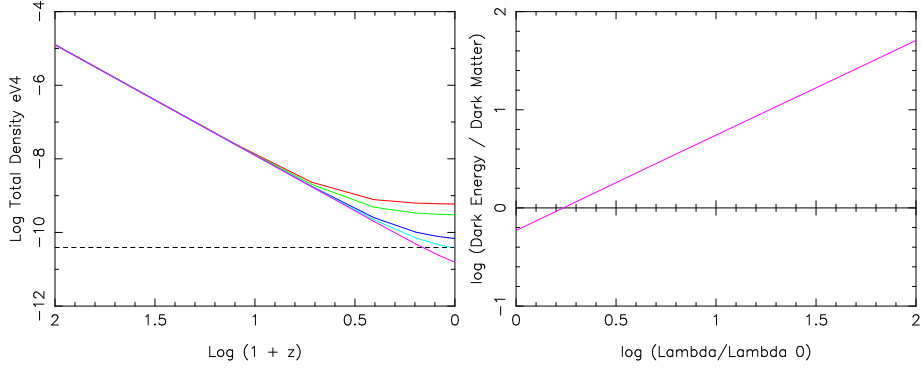

\begin{center}
\psfig{figure=quintotdens.eps,angle=-90,width=6cm}
\psfig{figure=quindmrel.eps,angle=-90,width=6cm}
\caption{Left: Evolution of total density with redshift for 
${\Gamma}_0 \equiv 
{\Gamma}_q/\Gamma = 10^{-16}$ (magenta) (see text for details), 
$5 {\Gamma}_0$ (cyan), $10 {\Gamma}_0$ (blue), $50 {\Gamma}_0$ 
(green), $100 {\Gamma}_0$ (red). Dash line is the observed value of the 
dark energy. $m_q = 10^{-6} eV$, $\lambda = 10^{-20}$. Right: 
Relative density of dark energy and CDM as a function of 
${\Gamma}_q/\Gamma$. The x-axis is normalized to ${\Gamma}_0 \equiv 
{\Gamma}_q/\Gamma = 10^{-16}$.
\label {fig:quintotdens}}
\end{center}
\end{figure}

\begin{figure}[h]
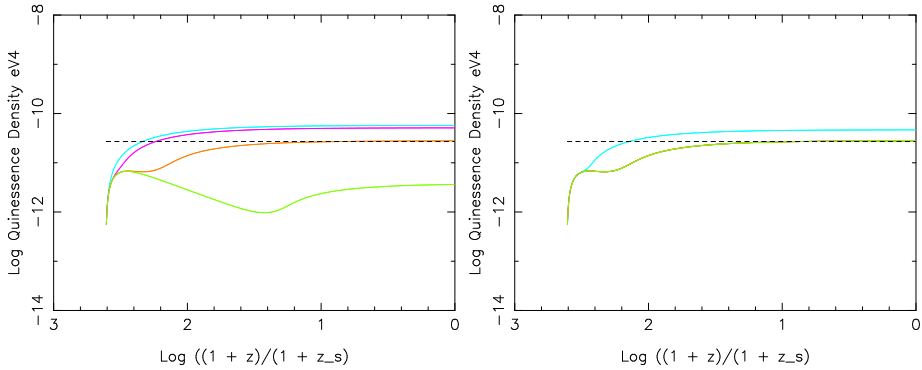

\begin{center}
\psfig{figure=quindenscompmass.eps,angle=-90,width=6cm}
\psfig{figure=quindenscomplambda.eps,angle=-90,width=6cm}
\caption{Quintessence energy density for: Left, 
$m_q = 10^{-3} eV$ (cyan), $m_q = 10^{-5} eV$ (magenta), $m_q = 10^{-6} eV$ 
(red) and $m_q = 10^{-8} eV$ (green), $\lambda = 10^{-20}$; Right, 
$\lambda = 10^{-10}$ (cyan), $\lambda = 10^{-15}$, $\lambda = 10^{-20}$ 
and $\lambda = 10^{-25}$ (green), $m_q = 10^{-6} eV$. The difference between 
quintessence density for the last 3 values of $\lambda$ is smaller than 
the resolution of 
the plot. Dash line is the observed energy density of the dark energy.
\label {fig:quinmass}}
\end{center}
\end{figure}

Fig.\ref{fig:quincontrib} shows the evolution in the contribution of 
different 
terms of the Lagrangian (\ref{lagrange}) to the total energy of $\phiq$. 
Very soon after beginning of production of quintessence field the potential 
takes over the kinetic energy and the latter begins to decrease. The 
relative contribution of each term and their time of dominance as this 
figure demonstrates, depends on the parameters specially $m_q$ and 
$\lambda$. Another conclusion from this plot is that changing these 
parameters by orders of magnitude does not change the general behavior of 
the model significantly and for a large part of the parameter space the 
final density of quintessence energy is close to the observed value. This 
can also be seen in Fig.\ref{fig:quintotdens} and Fig.\ref{fig:quinmass} 
where the evolution of quintessence energy is shown for various combination 
of parameters. 

\section {Perturbations} \label {sec.perturb}
Large and medium scale observations show that the dark energy is quite 
smooth and uncorrelated from the clumpy dark matter ~\cite {hubblevar}. 
If DE origin is the decay of the dark matter, the question arises whether 
it clumps around dark matter halos or has a large scale perturbation which 
is not observed in the present data. 
In this section we investigate the evolution of spatial perturbations in 
$\phiq$ and show that they decrease with time. Another interest in doing such 
exercise is to investigate any imprint of the model on the power spectrum 
of matter and the CMB anisotropy.

We use the synchronous gauge metric:
\be
ds^2 = dt^2 - a^2 (t) ({\delta}_{ij} - h_{ij}) dx^i dx^j \label {metric}
\ee
For small spatial fluctuations $\phiq (x,t) = \bar{\phiq}(t) + 
\delta \phiq (x,t)$ where from now on barred quantities are the homogeneous 
component of the field depending only on $t$. We define the same 
decomposition for other fields.

We consider only scalar metric fluctuations $h \equiv {\delta}^{ij} h_{ij}$ 
and neglect vector and tensor components. The Einstein equation gives 
following equation for the evolution of $h$:
\be
\frac {1}{2} \ddot {h} + \frac {\dot {a}}{a} \dot {h} = 4 \pi G (4 
\dot {\bar{\phiq}}\dot {\delta \phiq} - 2 \delta V (\phiq, {\rho}_x) + 
\delta {\rho}_x + \delta {\rho}_J + 3 \delta P_J) \label {deltheq}
\ee
where $\delta {\rho}_x$ is the fluctuation of $X$ particles density, 
$\delta {\rho}_J$ and $\delta P_J$ are respectively the collective density 
and pressure fluctuation of other fields. From the Lagrangian 
(\ref{lagrange}), the dynamic equation of $\phiq$ is:
\be
{\partial}_{\mu} (\sqrt {-g} g^{\mu\nu} {\partial}_{\nu}\phiq) + 
\sqrt {-g} V'(\phiq,\phix,J) = 0 \label {qgeq}
\ee
This equation and the energy momentum conservation determine the evolution 
of $\delta \phiq (x,t)$:
\bea
& & \dot {\bar {\phiq}}\biggl [ \ddot {\delta \phiq} + {\partial}_i
{\partial}^i (\delta \phiq) + V_q''(\bar {\phiq})\delta \phiq + 
2g \biggl (\frac {2\bar {\rho}_x}{{m_x}^2}\biggr ) \delta \phiq + 
3 \frac {\dot {a}}{a} \dot {\delta \phiq} \biggr ] + 
\frac{2g \bar {\phiq}}{{m_x}^2} \biggl [2 \frac {\dot 
{\rho}_x}{\bar {\rho}_x} \delta \phiq + \bar {\phiq} \frac {\dot {\delta 
{\rho}_x}}{\bar {\rho}_x}\biggr ] - \nonumber \\
 & & \hspace {1cm} \frac {\dot {a}}{a} 
\biggl [h \biggl (\frac {1}{2}\dot {\bar{\phiq}}^2 - V (\bar {\phiq})
\biggr ) - 6 \biggl ({V_q}' \delta \phiq + \frac{2g \bar {\phiq} 
\bar {\rho}_x}{{m_x}^2} (2 \delta \phiq + \bar {\phiq} \frac {\delta {\rho}_x}
{\bar {\rho}_x} )\biggr )\biggr ] - 
\frac {\dot {h}}{2}\dot {\bar{\phiq}}^2 = {\Gamma}_q (\delta {\rho}_x - 
\frac {\dot {\delta \phiq}}{\dot {\bar{\phiq}}} \bar {\rho}_x) 
\label {phidoteq} \\
\eea
Like homogeneous case, we assume that SDM behaves like a pressure-less fluid:
\be
{T_x}^{00} = \bar {\rho}_x + \delta {\rho}_x \quad \quad {T_x}^{0i} = 
\bar {\rho}_x \delta {u_x}^i \quad \quad {T_x}^{ij} = \mathcal {O} 
({\delta}^2) \thickapprox 0 \label {tx}
\ee
where $\delta {u_x}^i$ is the velocity of SDM fluctuations with respect to 
homogeneous Hubble flow. Interaction terms are explicitly included 
in the energy-momentum conservation equation:
\be
{\partial}_0 \biggl (\frac {\delta {\rho}_x}{\bar {\rho}_x} \biggr ) + 
{\partial}_i (\delta {u_x}^i) - \frac {\dot {h}}{2} = - {\pi}^4 g^2 \biggl 
(\frac {3 \delta {\rho}_x}{{m_x}^3} - \frac {2 \bar {{\rho}_q}' 
\delta {{\rho}_q}'}{{m_q}^3 \bar {\rho}_x} - \frac {\bar {{\rho}_q}'^2 
\delta {{\rho}_x}}{{m_q}^3 \bar {{\rho}_x}^2}\biggr ) \label {deltrhoxeq}
\ee
The effect of interactions in the right hand side of (\ref {deltrhoxeq}) is 
however very small, first because $X$ particles mass is very large and then 
because only high energy $\phiq$ particles contribute to this term and their 
energy decreases with expansion of the Universe much faster than the SDM. 
The evolution of matter fluctuations is then practically the same as 
the standard $\Lambda$CDM case.

Using the conservation relation for other components of the energy-momentum 
in the limit when $\dot {\bar {\phiq}} \rightarrow 0$, we find the following 
relation between spatial fluctuation of $\delta \phiq$ and $\delta {u_x}^i$:
\be
- V' (\bar{\phiq}, \bar {\rho}_x) {\partial}^i (\delta 
\phiq) = {\Gamma}_q \bar {\rho}_x \delta {u_x}^i \label {phidotueq}
\ee
Equation (\ref {phidoteq}) has a meaningful limit when $\dot {\bar {\phiq}} 
\rightarrow 0$ only if $\dot {\delta \phiq} \rightarrow 0$. On the other 
hand, (\ref {phidotueq}) shows that the divergence of quintessence field 
fluctuations ${\partial}^i \delta \phiq$ follows the velocity 
dispersion of the dark matter with opposite direction. Their amplitude 
however is largely reduced due to the very small decay width ${\Gamma}_q$. 
In addition, with the expansion of 
the Universe, $V' (\bar{\phiq}, \bar {\rho}_x)$ varies only very slightly -
 just the interaction between SDM and $\phiq$ will change. In contrast,  
$\bar {\rho}_x$ decreases by a factor of $a^{-3}(t)$ and even gradual 
increase of the dark matter clumping and therefore the velocity dispersion 
$\delta {u_x}^i$ ~\cite {hubblevar} can not eliminate the effect of 
decreasing density. We conclude that the spatial variation of 
$\phiq$ is very small from the beginning and is practically unobservable.

\section {Closing Remarks}
Since the original works on the production of superheavy particles after 
inflation ~\cite {grapro}, a number of investigations ~\cite {wimprod} have 
demonstrated that even with a reheating temperature as low as few MeV the 
production of superheavy particles is possible. We don't discuss here the 
particle physics candidates for $\phiq$, but for the sake of completeness 
we just mention that axion like particles are needed or at least can exist 
in large number of particle physics models (see ~\cite {axion} for some 
examples). The fact that $\phiq$ does not need to have very special 
potential is one of the advantages of this model with respect to others and 
opens the way to a larger number of particle physics models as candidate 
for the quintessence field.

One of the arguments which is usually raised in the literature against a 
decaying dark 
matter is the observational constraints on the high energy gamma-ray and 
neutrino background. In ~\cite {wimpzhouri} it has been shown that if 
$m_x \gtrsim 10^{22} eV$ and its lifetime $\tau \gtrsim 5 {\tau}_0$, and if 
simulations correctly take into account the energy dissipation of the high 
energy remnants, present observational limits are larger than expected 
flux from a decaying UHDM. Consequently the model is consistent with the 
available data. 

The same fact is applied to 
the CMB and its anisotropy. The expected CMB distortion is of order 
$10^{-8}$, much smaller than sensitivity of present and near future 
measurements. As for the expected anisotropy in the arrival direction of 
UHECRs, the data is yet too scarce to give any conclusive answer. In the 
next few years the Auger Observatory ~\cite {auger} is able to test 
top-down models for UHECRs which is one of the principal motivation for the 
quintessence model proposed here.

Although the limit on the amount of the hot DM can not constrain this model, 
a better understanding of its contribution to the total density and its 
content can help to understand the physics and the nature of SDM if it 
exists. 

Evidently the observation of $w_q$ and its cosmological evolution is 
crucial for any model of dark energy. Observation of small anisotropy in the 
DE density and its correlation with matter anisotropy also can 
be used as signature of relation/interaction between DM and DE.

For the range of expected masses for $\phiq$ in this model, the high energy 
component of quintessence field is yet relativistic. As we have discussed in 
Sec.\ref {sec.model}, the production of this component from annihilation has 
been stopped very early in the history of the Universe and the contribution 
from decay of $X$ is much smaller than the limits on the amount of Hot Dark 
Matter (as it has been shown in ~\cite {wimpzhouri} for hot SM remnants). 
The small coupling of $\phiq$ with SM particles also suppresses the 
probability of its direct detection. However, the detection of an 
axion-like particle e.g. the QCD axion can be a positive sign for the 
possibility of existence of $\phiq$-like particles in the Nature. 

\section*{Appendix}
Here we calculate the right hand side of the Boltzmann equation at lowest 
order of $g$ coupling constant for annihilation and reproduction of $X$ 
particles.

The Boltzmann equation for $X$ particles is the following:
\be
p^{\mu}{\partial}_{\mu} f^{(X)}(x,p) - {\Gamma}^{\mu}_{\nu\rho} p^{\nu}
p^{\rho} \frac {\partial f^{(X)}}{\partial
p^{\mu}} \equiv L[f] = -({\mathcal A}(x,p) + {\mathcal B}(x,p)) f^{(X)}(x,p) +
{\mathcal C}(x,p) \label {bolt}
\ee
\bea
{\mathcal A}(x,p) & = & {\Gamma} p^{\mu} u_{\mu}.\label {idec}\\
{\mathcal B}(x,p) & = & \frac {1}{(2\pi)^3 g_x} \sum_i \int d\bar p_x
f^{(i)}(x,p_x) A {\sigma}_{xi} \quad\quad i = X,\phiq \label {absint}\\
{\mathcal C}(x,p) & = & \frac {1}{(2\pi)^6 g_x}\int d\bar p_q
d\bar p_q f^{(q)}(x,p_q)f^{(q)}(x,p'_q) A \frac {d{\sigma}_{\phiq + \phiq
\rightarrow X + X}}{d\bar p} \label {proint}\\
A & = &\sqrt {(p_1.p_2)^2 - {m_1}^2{m_2}^2} \label {adef}
\eea

The function $f^{(X)} (x,p)$ is the distribution of $X$ particles. The terms 
${\mathcal A}$, ${\mathcal B}$ and ${\mathcal C}$ are respectively the decay, 
the annihilation (self or in interaction with other species) and production 
rates. We assume that the interaction of $X$ with other fields 
except $\phiq$ is negligible. According to Lagrangian (\ref {lagrange}) with 
$n = 2$ and $m = 2$ the lowest Feynman diagrams contributing to annihilation 
and production are:

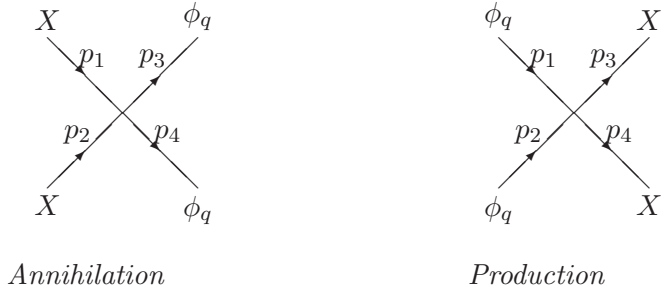
\begin {picture}(90,50)(-50,-20)
\linethickness{1pt}
\put(0,0){\vector(1,1){5}}
\put(5,5){\line(1,1){5}}
\put(10,10){\vector(1,1){5}}
\put(15,15){\line(1,1){5}}
\put(0,20){\vector(1,-1){5}}
\put(5,15){\line(1,-1){5}}
\put(10,10){\vector(1,-1){5}}
\put(15,5){\line(1,-1){5}}
\put(0,-1){\makebox(0,0)[t]{$X$}}
\put(4,6){\makebox(0,0)[b]{$p_2$}}
\put(0,21){\makebox(0,0)[b]{$X$}}
\put(6,16){\makebox(0,0)[b]{$p_1$}}
\put(20,-1){\makebox(0,0)[t]{$\phiq$}}
\put(16,6){\makebox(0,0)[b]{$p_4$}}
\put(20,21){\makebox(0,0)[b]{$\phiq$}}
\put(14,16){\makebox(0,0)[b]{$p_3$}}
\put(5,-10){\makebox(0,0)[t]{\it{Annihilation}}}

\put(60,0){\vector(1,1){5}}
\put(65,5){\line(1,1){5}}
\put(70,10){\vector(1,1){5}}
\put(75,15){\line(1,1){5}}
\put(60,20){\vector(1,-1){5}}
\put(65,15){\line(1,-1){5}}
\put(70,10){\vector(1,-1){5}}
\put(75,5){\line(1,-1){5}}
\put(60,-1){\makebox(0,0)[t]{$\phiq$}}
\put(64,6){\makebox(0,0)[b]{$p_2$}}
\put(60,21){\makebox(0,0)[b]{$\phiq$}}
\put(66,16){\makebox(0,0)[b]{$p_1$}}
\put(80,-1){\makebox(0,0)[t]{$X$}}
\put(76,6){\makebox(0,0)[b]{$p_4$}}
\put(80,21){\makebox(0,0)[b]{$X$}}
\put(74,16){\makebox(0,0)[b]{$p_3$}}
\put(65,-10){\makebox(0,0)[t]{\it{Production}}}
\end {picture}

The $S$ matrix for these diagrams is very simple:
\be
S = \frac {-i (2\pi)^4 g {\delta}^{(4)}(\sum_i p_i)}{\prod_i 2 {p_i}^0} 
\label {smat}
\ee
and the differential cross-section:
\be
d\sigma = \frac {(2\pi)^{10} g^2 {\delta}^{(4)}(\sum_i p_i)}
{16 \sqrt {(p_1.p_2)^2 - {m_1}^2{m_2}^2}} d\bar {p_3}d\bar {p_4} \quad \quad 
d\bar {p_i} \equiv \frac {d^3p_i}{(2\pi)^3 g_i {p_i}^0} \label {dsig}\\
\ee
where $g_i$ is the number of internal degrees of freedom. 
Here we assume that $g_x = g_q = 1$. Using the relation:
\be
\biggl [\int p^{{\mu}}p^{{\mu}^1} \ldots p^{{\mu}^n} f (x,p) d\bar{p} 
\biggr ]_{;\mu} = \int p^{{\mu}^1} \ldots p^{{\mu}^n} L[f] (x,p) d\bar{p} 
\label {ldiff}
\ee
and the definition of energy-momentum tensor $T^{\mu\nu}$ and number 
density of particles $n^{\mu}$, one obtains:
\be
{T^{\mu\nu}}_{;\nu} = -\Gamma T^{\mu\nu} u_{\nu} - {\pi}^4 g^2 
\biggl ({n_x}^{\mu}\sum_i \int d\bar {p}_2 f^{(i)}(x,p_2) - \int d\bar {p}_1 
d\bar {p}_2 {p_1}^{\mu} f^{(q)}(x,p_1) f^{(q)}(x,p_2) \theta ({p_1}^0 + 
{p_2}^0 - 2 m_x) \biggr ) \label {momcons}
\ee
Both $f^{(X)}$ and $f^{(q)}$ have a large peak around the energies close to 
the mass of $X$. Therefore:
\be
\int d\bar {p} f^{(i)}(x,p) \approx \frac {{n_i}^{\nu} \bar {u}_{\nu}}{m_i} 
\quad\quad i = X,q \label {intf}
\ee
In the case of $\phiq$ the density $n_q$ is only the density of particles 
with an average energy larger than $m_x$. Finally from (\ref {momcons}) one 
can obtain the evolution equation of ${\rho}_x$ in a homogeneous cosmology 
i.e. equation (\ref {xeq}) in Sec.\ref{sec.model}.

\end{document}